\documentclass{article}
\usepackage[centertags]{amsmath}
\usepackage{amssymb,texdraw,a4wide,epsfig}
\numberwithin{equation}{section}
\font\bigsc=cmcsc10 scaled \magstep 1
\font\bigrm=cmr12
\font\bbigbf=cmbx12 scaled \magstep 1
\font\bigbf=cmbx12
\font\filt=msbm10
\def\filR{\hbox{\filt R}}
\def\filN{\hbox{\filt N}}
\def\bv#1{\hbox{$\mathbf{#1}$}}
\def\sgn{\mathop{\rm sgn}\nolimits}
\def\vth{\varphi}
\def\vph{\vartheta}
\def\vpi{\varpi}
\def\vrh{\varrho}
\def\psio{\psi_\circ}
\def\eps{\epsilon}
\def\vsgo{\varsigma_0}
\def\rs{r_\circ}
\def\ac{a_\circ}
\def\as{A_\circ}
\def\vs{V_\circ}
\def\dps{\displaystyle}
\def\pqa{\Big[\hskip-.28em\Big[}
\def\pqc{\Big]\hskip-.28em\Big]}
\begin{document}
\begin{titlepage}
\null \vfil

\centerline{\bbigbf Impermeability effects in three-dimensional
vesicles}

\vfil\vfil

\centerline{\bigsc Paolo Biscari$^{(1)}$, Silvia Maria
Canevese$^{(2)}$, {\bigrm and} Gaetano Napoli$^{(3)}$}
\bigskip
\noindent\small
\renewcommand{\labelenumi}{$^{(\theenumi)}$}
\begin{enumerate}
\item Dipartimento di Matematica, Politecnico di Milano.
Piazza Leonardo da Vinci 32 -- 20133 Milano (Italy) and Istituto
Nazionale di Fisica della Materia. Via Ferrata 1 -- 27100 Pavia
(Italy). \\ E-mail: paolo.biscari@polimi.it
\item Dipartimento di Elettronica e Informazione, Politecnico di
Milano. Via Ponzio 34 -- 20133 Milano (Italy). E-mail:
canevese@elet.polimi.it
\item Dipartimento di Matematica, Politecnico di Milano.
Piazza Leonardo da Vinci 32 -- 20133 Milano (Italy). E-mail:
gaetano.napoli@mate.polimi.it
\end{enumerate}

\normalsize

\vfil\vfil

\noindent 2003 PACS: \ $\vtop{\hsize 11 cm \noindent 87.16.Dg
(Membranes, bilayers, and vesicles)\hfil\break 87.15.Kg (Molecular
interactions; membrane-protein interactions)}$

\vfil

\noindent 2000 MSC: \ \ $\vtop{\hsize 9 cm \noindent 74L15
(Biomechanical solid mechanics) \hfil\break 74K15 (Membranes)
\hfil\break 74B15 (Equations linearized about a deformed state)}$

\vfil

\noindent Submitted to: {\sl Journal of Physics A: Mathematical
and General} (2004)

\vfil\vfil

\begin{abstract}
We analyse the effects that the impermeability constraint induces
on the equilibrium shapes of a three-dimensional vesicle hosting a
rigid inclusion. A given alteration of the inclusion and/or
vesicle parameters leads to shape modifications of different
orders of magnitude, when applied to permeable or impermeable
vesicles. Moreover, the enclosed-volume constraint wrecks the
uniqueness of stationary equilibrium shapes, and gives rise to
pear-shaped or stomatocyte-like vesicles.
\end{abstract}

\vfil\vfil

\noindent Date: April 28, 2004

\vfil\vfil
\end{titlepage}

\section{Introduction}

Lipid proteins embedded in biological membranes strongly influence
both the geometric and the elastic properties of the hosting
membranes. The rigid inclusions are able to induce vesicle budding
\cite{96scor,01huzi,02husc}, while the interplay between the
protein-membrane interaction and the spontaneous curvature may
yield a loss of regular equilibrium configurations \cite{01biro}.
Moreover, the elasticity of the hosting vesicle induces a
membrane-mediated interaction that has been widely studied both
experimentally \cite{99kora} and theoretically, in the cases of
planar \cite{02bibi,02bibiro}, quasi-planar
\cite{93gobr,96gogo,98weko}, and quasi-spherical vesicles
\cite{98dofo}.

In this paper we focus attention on the equilibrium shapes of a
three-dimensional vesicle hosting a rigid inclusion, and in
particular on the effects of the permeability properties of the
vesicle. In fact, a given slight perturbation applied to a
quasi-spherical vesicle may induce quite different changes on the
resulting equilibrium shape, depending on whether the enclosed
volume of the vesicle is constrained or not. More precisely, an
$O(\epsilon)$ relative perturbation of the external parameters
induces an equivalent $O(\epsilon)$ modification in the shape
function of a permeable vesicle, but a quite stronger
$O(\sqrt{\epsilon}\,)$ relative perturbation if the enclosed
volume is kept fixed. Furthermore, the volume constraint yields a
multiplicity of stationary equilibrium shapes, and it leads to
abandon the spherical shape towards either pear-shaped or
stomatocyte-like vesicles.

The volume enclosed by the vesicle may be constrained or not
depending on both the chemical properties of the aqueous solution
which surrounds the vesicle, and the time scales in which the
(meta)equilibrium shapes are observed \cite{91se,96juli}. When the
water is essentially free of molecules that cannot permeate the
bilayer membrane, no volume constraint stands. On the contrary,
when some of the molecules in the solution are unable to permeate
the bilayer, the resulting osmotic pressure gives rise to an
enclosed-volume constraint. However, even in this latter case, on
long time scales water molecules succeed in permeating the
membrane. Eventually, the vesicle reaches its true equilibrium
shape, which minimizes the free energy with respect to the
enclosed volume.

Throughout our development, we will model proteins as rigid
inclusions, and we will assume that the protein-membrane
interaction simply fixes the contact angle, \emph{i.e.\/} the
angle the vesicle normal determines with the inclusion plane
\cite{93dapi,93gobr,98dasa}. However, only slight modifications
need to be applied to our results to take into account
interactions which determine the contact curvature instead of the
contact angle \cite{96palu,99dofo,02dofo}. More drastic changes,
though substantially the same mathematical setting, requires the
weak anchoring case \cite{98dasa,00nian,01biro}, in which the
inclusion-vesicle interaction provides an additional term in the
free-energy functional, instead of a fixed boundary condition.

We describe vesicle elasticity through Helfrich's (spontaneous
curvature) model \cite{73he,76dehe}. The free-energy functional is
then
\begin{equation}
\mathcal{F}\big[\Sigma\big]:=\kappa\int_\Sigma\big(H-\sigma_0
\big)^2da\;,
\label{fren}
\end{equation}
where $\Sigma$ is a closed surface describing the vesicle shape,
$H$ denotes the mean curvature along $\Sigma$, $\kappa$ is the
bending energy, and $\sigma_0$ the spontaneous curvature. In the
minimizing procedure at fixed area (and possibly fixed volume) we
replace (\ref{fren}) with the effective free-energy functional
\begin{equation}
\mathcal{F}_{\rm eff}
\big[\Sigma\big]:=\kappa\int_\Sigma\big(H-\sigma_0\big)^2da+\lambda
\,\big({\rm Area}\,(\Sigma)-A\big)+\pqa\mu \,\big({\rm Vol}\,
(\Sigma)-V\big)\;\pqc\;.
\label{flagr}
\end{equation}
The Lagrange multipliers $\lambda$, $\mu$ have the physical
meaning of surface tension and pressure difference, and the
brackets are there to remind that the volume constraint will not
be always applied. In (\ref{flagr}) the area constraint has been
inserted as a global, instead of a local constraint. We recall
that, in the absence of external forces, both choices are
equivalent \cite{87zhhe}.

The Euler-Lagrange equation associated to the functional
(\ref{flagr}) is the \emph{shape equation\/} \cite{87zhhe,89zhhe}:
\begin{equation}
\kappa\left[\Delta_{\rm
s}H+2H\left(H^2-K\right)+2\sigma_0\,K-2\sigma_0^2H\right]-2\lambda
\, H-\pqa \mu\pqc =0\;,
\label{eulag}
\end{equation}
where $\Delta_{\rm s}$, the tangential divergence of the
tangential gradient, is the Laplace-Beltrami operator on $\Sigma$,
and $K$ denotes the Gaussian curvature along $\Sigma$.

The plan of the paper is as follows. In the next section, we
introduce the surface parameterization and we derive the
conditions satisfied by spherical and quasi-spherical shapes. In
Section 3 we analyse the vesicle shapes obtained in the presence
of the area constraint alone (\emph{i.e.\/}, the long-time
equilibrium vesicle shapes). Section 4 is devoted to the peculiar
role played by the volume constraint. In Section 5 we review and
discuss our results.

\section{The model}

\begin{figure}
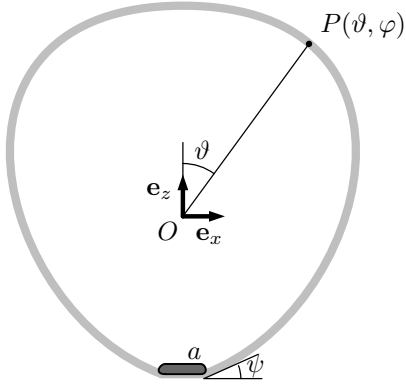

\input figure1.tex
\caption{Cross-section of a three-dimensional vesicle embedding a
conical inclusion of base radius $a$. The tangent to the
generating curve at the contact points determines a fixed angle
$\psi$ with the inclusion plane.}
\label{setting}
\end{figure}

Let us consider a vesicle which embeds an inclusion, that we model
as a symmetric conical frustum of negligible height, base radius
$a$, and apex angle $\psi$. The inclusion-vesicle interaction
fixes the angle between the vesicle normal and the inclusion plane
at the contact points to be equal to
$\left(\frac{\pi}{2}-\psi\right)$. Figure \ref{setting}
illustrates the geometric setup of the model. For a more detailed
description of the inclusion-vesicle interactions and their
modeling we refer the reader to the paper by Biscari and Rosso
\cite{01biro}.

We restrict our analysis to axisymmetric vesicle shapes, and
parameterize them in spherical coordinates centered in a fixed
point $O$, which lies in the inclusion symmetry axis $z$ (we
assume that this is the symmetry axis of the vesicle too, since
the inclusion does not upset the cylindrical symmetry):
$$
P(\vph,\vth)-O=r(\vph)\sin\vph\cos\vth
\,\bv{e}_x+r(\vph)\sin\vph\sin\vth \,\bv{e}_y+r(\vph)\cos\vph
\,\bv{e}_z\;.
$$
The unit vectors $\{\bv{e}_x,\bv{e}_y,\bv{e}_z\}$ form an
orthogonal basis, with $\bv{e}_z$ parallel to the inclusion axis;
($\vph,\vth$) are the polar and azimuthal angles, respectively.
The area element, the Laplace-Beltrami operator, the mean
curvature, and the Gaussian curvature along $\Sigma$ are given by:
\begin{align*}
da&=\sqrt{g}\;d\vph\,d\vth\;,\qquad \Delta_{\rm
s}=\frac{1}{\sqrt{g}}\left[\partial_\vph
\left(\frac{r\sin\vph}{\sqrt{r^2+r^{\prime
2}}}\,\partial_\vph\right)+\partial_\vth
\left(\frac{\sqrt{r^2+r^{\prime
2}}}{r\sin\vph}\,\partial_\vth\right)\right],\\
\noalign{\medskip}
H&=\frac{2r^3+3r r^{\prime 2}-r^{\prime 3}\cot\vph -r^2 (
r'\cot\vph +r'')}{2\,r\left(r^2+r^{\prime
2}\right)^{3/2}}\;,\qquad {\rm and}\\
K&=\frac{(r\sin\vph-r'\cos\vph)\left(r^2+2r^{\prime 2}-r
\,r''\right)}{r\,\sin\vph\left(r^2+r^{\prime 2}\right)^2}\;,
\end{align*}
where a prime denotes differentiation with respect to the polar
angle, and $g:=\big(r^2+r^{\prime 2}\big)r^2\sin^2\vph\;$. The
spherical parameterization transforms the effective free energy
(\ref{flagr}) as follows:
$$
\mathcal{F}_{\rm eff} \big[r\big]=2\pi\int_0^{\vph_{\rm f}}
\left(\kappa\,\big(H[r]-\sigma_0\big)^2+\lambda\right)\,r\,
\sqrt{r^2+r^{\prime 2}}\,\sin\vph\,d\vph+\left[\hskip-.28em
\left[\frac{2\pi}{3} \mu\int_0^{\vph_{\rm f}} r^3(\vph)\sin\vph
\,d\vph\right]\hskip-.28em\right].
$$
The integration limit $\vph_{\rm f}$ depends on position of the
origin $O$. We choose to fix $O$ at a distance $\Delta
z=(a\cot\psi)$ above the inclusion, which yields $\vph_{\rm
f}=(\pi-\psi)$. This choice simplifies the attachment condition on
the inclusion and the constraint on the direction of the contact
normal, which become:
\begin{equation}
r\left(\vph_{\rm f}\right)=\frac{a}{\sin\psi} \qquad{\rm
and}\qquad r'(\vph_{\rm f})=0\;.
\label{contcond}
\end{equation}
The vesicle is free on its top ($\vph=0$). The boundary conditions
therein follow from regularity requirements on the vesicle shape:
\begin{equation}
\lim_{\vph\to 0^+} r'(\vph)=0\qquad{\rm and}\qquad \lim_{\vph\to
0^+} r'''(\vph)=0\;.
\label{freecond}
\end{equation}

\subsection{Internal actions}

Let us consider a subsurface $\Sigma'\subseteq \Sigma$, and let
\bv{n} and \bv{t} respectively denote the normal to $\Sigma$ at a
point $P\in\partial\Sigma'$, and the tangent to the curve
$\partial\Sigma'$. Furthermore, let $\bv{k}:=\bv{n}\wedge\bv{t}$
denote the direction in the tangent plane at $P$ pointing outwards
with respect to $\Sigma'$. The internal actions at $P$ consist in
a distributed force \bv{f} and a distributed torque \bv{m}, whose
densities per unit length of $\partial \Sigma'$ are
$$
\bv{f}=\left[\kappa\,\big(H-\sigma_0\big)^2+\lambda\right]\,\bv{k}
-\kappa\,\frac{\partial H}{\partial k}\,\bv{n}\qquad {\rm
and}\qquad \bv{m}=\kappa\,\big(H-\sigma_0\big)\,\bv{n}\;.
$$
The above equations generalize to three-dimensional vesicles the
internal actions derived in \cite{02bibiro} in the two-dimensional
case. They show that the surface tension $\lambda$ may become
negative without giving rise to the collapse of the vesicle,
provided that $\bv{f}\cdot\bv{k}= \big[\kappa\,(H-
\sigma_0)^2+\lambda\big]$ remains non-negative all along the
vesicle. Furthermore, they provide an alternative way of deriving
the free-boundary conditions (\ref{freecond}). In fact, these
conditions are equivalent to the vanishing of the internal force
and torque acting on an infinitesimal cap which surrounds the
vesicle top.

\subsection{Spherical shapes}

The equilibrium shape of a vesicle is a sphere of radius $\rs$,
centered in $O$, whenever the base radius $\ac$, the apex angle
$\psio$, and the vesicle area $\as$ satisfy
\begin{equation}
\ac=\rs\sin\psio\qquad{\rm and}\qquad
\as=4\pi\rs^2\cos^2\frac{\psio}{2} \;.
\label{c1sph}
\end{equation}
If we eliminate $\rs$ from equations (\ref{c1sph}$)_1$ and
(\ref{c1sph}$)_2$, we obtain:
$$
\pi \ac^2=\as \sin^2\frac{\psio}{2}\;.
$$
If, in addition, the vesicle is impermeable, the enclosed volume
$\vs$ must match
$$
\vs=\frac{4}{3}\pi\rs^3\cos^2\frac{\psio}{2}+\frac{1}{3} \pi
\ac^2\rs\cos\psio=\frac{\pi \ac^3}{6}\,
\frac{\cos\frac{\psio}{2}}{\sin^3\frac{\psio}{2}} \,
\big(2-\cos\psio\big).
$$
In particular, area, enclosed volume, and apex angle must obey:
\begin{equation}
\upsilon_\circ:=36 \pi
\frac{\vs^2}{\as^3}=(2-\cos\psio)^2\,\cos^2\frac{\psio}{2}\;.
\label{c2sph}
\end{equation}
We remark that for any value in $\upsilon_\circ\in[1,2]$ (the end
cases corresponding to the cases of a sphere and a half-sphere),
there is exactly one value of $\psio\in\big[0,\frac{\pi}{2}\big]$
that satisfies (\ref{c2sph}).

\subsection{Quasi-spherical shapes}

We now assume that some of the control parameters $a$, $\psi$, $A$
(and possibly $V$) are slightly perturbed with respect to their
values satisfying (\ref{c1sph}) and (\ref{c2sph}). In this case,
we look for solutions of the shape equation (\ref{eulag}) that
represent a perturbation of a sphere:
\begin{equation}
r(\vph)=\rs\, \big(1+\eps\,\vrh_1(\vph)+o(\eps)\big)\;.
\label{exp1}
\end{equation}
Consistently, we also expand the Lagrange multipliers by
perturbing their ``spherical'' values:
\begin{equation}
\lambda=\frac{\kappa}{\rs^2}\,\big(\Lambda_0 +\eps\,\Lambda_1
+o(\eps)\big)\qquad {\rm and}\qquad
\mu=\frac{2\kappa}{\rs^3}\,\big(\eta_0
+\eps\,\eta_1+o(\eps)\big)\;.
\label{lamu1}
\end{equation}
In (\ref{lamu1}), the normalizing factors $\kappa$ and $\rs$ have
been inserted in order to proceed with the dimensionless
quantities $\Lambda_i$ and $\eta_i$. Furthermore, in order to make
the whole shape equation dimensionless, we define the reduced
spontaneous curvature as
\begin{equation}
\vsgo:=\sigma_0\,\rs\;.
\label{dsig}
\end{equation}

If we insert (\ref{exp1}), (\ref{lamu1}) and (\ref{dsig}) in
(\ref{eulag}), we derive at $O(1)$ the condition
\begin{equation}
\Lambda_0+\eta_0=\vsgo\,(1-\vsgo)\;,
\label{ord0}
\end{equation}
linking the spherical values of the Lagrange multipliers.

When we push the expansion to $O(\eps)$, we obtain a fourth-order
linear differential equation for $\vrh_1$. If we further introduce
the variable $s:=\cos\vph$, and perform the substitution
$\vrh_1(\vph)=\vpi_1(\cos\vph)$, the so-obtained differential
equation reads as:
\begin{equation}
\left(1-s^2\right)^2\!\vpi^{(4)}\!-8s\left(1-s^2\right)\vpi^{(3)}+
\left[12s^2-4+g_1\left(1-s^2\right)\right]\vpi^{(2)}\!-
2g_1s\vpi^{(1)}+g_0 \vpi=-4(\Lambda_1+\eta_1),
\label{lin1}
\end{equation}
where
$$
g_0=2g_1:=4\big(\vsgo(2-\vsgo)-\Lambda_0\big)\;,
$$
and the superscripts denote differentiation with respect to $s$.
Equation (\ref{lin1}) is an inhomogeneous fourth-order Legendre
differential equation. Its general solution can be expressed in
terms of Legendre functions of the first and second kind as:
$$
\vpi(s)=-\frac{4(\Lambda_1+\eta_1)}{g_0}+C_1\,P_{\nu_+}(s)+
C_2\,Q_{\nu_+}(s)+C_3\,P_{\nu_-}(s)+C_4\,Q_{\nu_-}(s)\;,
$$
where the orders of the Legendre functions are given by
\begin{equation}
\nu_{\pm}=-\frac{1}{2}+\frac{1}{2}\,\sqrt{5+2g_1\pm
2\sqrt{(g_1+2)^2-4g_0}}\;.\label{nupm}
\end{equation}
If we replace $g_0=2g_1$ in (\ref{nupm}), we obtain \quad $\dps{
\nu_\pm=-\frac{1}{2}+\frac{1}{2}\,\sqrt{5+2g_1\pm
2\left|g_1-2\right|\,}\;}$,\quad so that
$$
\big\{\nu_+\,,\,\nu_-\big\}=\left\{\,1\,,\,\frac{1}{2}
\left(\sqrt{1+4g_1}-1\right)\,\right\}\;.
$$
If we further introduce the notation
\begin{equation}
\nu:=\frac{1}{2}\left(\sqrt{1+4g_1}-1\right)
=\frac{1}{2}\left(\sqrt{1+8\vsgo(2-\vsgo)-8\Lambda_0\,}-1\right)\;,
\label{nu}
\end{equation}
and we finally consider that $P_1(s)=s\,$, we arrive at the
general solution of the linearized shape equation:
\begin{equation}
\vrh_1(\vph)=-\frac{4(\Lambda_1+\eta_1)}{g_0}+C_1\,\cos\vph+
C_2\,Q_1(\cos\vph)+C_3\,P_\nu(\cos\vph)+C_4\,Q_\nu(\cos\vph)\;.
\label{form1}
\end{equation}
Some further investigation will turn out to be necessary in the
particular cases $\nu=0$ ({\sl i.e.\/} $g_1=0$) and $\nu=1$ ({\sl
i.e.\/} $g_1=2$). We postpone the analysis of these cases to the
sections below.

In the following, we will analyse and compare the perturbed
equilibrium vesicle shapes of permeable and impermeable vesicles.
The derivation below works when any or even all of the parameters
$a$, $\psi$, $A$, or $V$ are varied with respect to their
spherical values. However, and only in order to shorten our
presentation, we will henceforth restrict our development to the
case in which only the area, and possibly the enclosed volume, are
varied with respect to the values satisfying (\ref{c1sph}) and
(\ref{c2sph}), while the inclusion parameters are kept unchanged.
These are the easiest perturbations to be implemented
experimentally: for example, an area variation in a biological
membrane may be simply induced by adding extra lipid molecules to
the membrane bilayer.

\section{Permeable vesicles\label{area}}

When the vesicle is inextensible but permeable, no volume
constraint stands. For all practical purposes, this is equivalent
to assume that the Lagrange multiplier $\mu$ vanishes identically.
When this is the case, condition (\ref{ord0}) reads as
$$
\Lambda_0=\vsgo(1-\vsgo)\;,
$$
which implies $g_0=2g_1=4\vsgo\;$, and
\begin{equation}
\nu=\frac{1}{2}\left(\sqrt{1+8\vsgo}-1\right)\;.
\label{nuso}
\end{equation}
The linear perturbation of the spherical shape becomes:
\begin{equation}
\vrh_1(\vph)=-\frac{\Lambda_1}{\vsgo}+C_1\,\cos\vph+
C_2\,Q_1(\cos\vph)+C_3\,P_\nu(\cos\vph)+C_4\,Q_\nu(\cos\vph).
\label{arlin}
\end{equation}
All Legendre functions of the second kind $Q_\nu(s)$ are singular
when $s\to1^-$. Thus, the free-boundary conditions
(\ref{freecond}) require $C_2=C_4=0$. The remaining parameters
$\Lambda_1$, $C_1$, and $C_3$ can be determined with the aid of
the contact conditions (\ref{contcond}) and the area constraint:
\begin{equation}
\vrh_1(\vph_{\rm f})=0\;,\qquad \vrh'_1(\vph_{\rm f})=0\;,\qquad
{\rm and}\qquad 4\pi\rs^2\int_0^{\vph_{\rm
f}}\vrh_1(\vph)\sin\vph\,d\vph=\Delta A\;,
\label{condfin}
\end{equation}
where $\Delta A$ is the area excess with respect to the spherical
value $\as$. With the aid of (\ref{8.5.3})-(\ref{intleg})
conditions (\ref{condfin}) yield
\begin{align*}
\Lambda_1&=\vsgo\,C_3\csc^2\psio\,\left[\nu\cos\psio
\,P_{\nu-1}(-\cos\psio)+\left(\sin^2\psio+\nu\cos^2\psio\right)
P_\nu(-\cos\psio)\right]\;,
\\
\noalign{\medskip}
C_1&=-\nu\,C_3\csc^2\psio\,\big[
P_{\nu-1}(-\cos\psio)+\cos\psio\,P_\nu(-\cos\psio)\big]\;, \qquad
{\rm and}\\
&\frac{C_3}{(\nu+1)(1-\cos\psio)}\; \left[ \frac{4 \big(
P_{\nu-1}(-\cos\psio)-P_\nu(-\cos\psio)\big)}
{1+\cos\psio}\right.\\
&-(\nu-1)(\nu\cos\psio+2)P_\nu(-\cos\psio)- (\nu^2+\nu+2)
P_{\nu-1}(-\cos\psio) \left]=\frac{\Delta A}{A_\circ}\;.\right.
\end{align*}

We have already announced that the differential equation
(\ref{lin1}) admits (\ref{arlin}) as its general solution only
when $\nu\not\in\{0,1\}$, that is when the spontaneous curvature
is neither null nor equal to the inverse of the unperturbed radius
$\rs$. We will now solve (\ref{lin1}) in these cases.

When $\vsgo=0$, the general solution of the homogeneous
differential equation associated to (\ref{lin1}) is still as in
(\ref{arlin}), with $\nu=0$. However, and since $P_0(s)\equiv 1$,
the particular solution of the equation is not a constant. By
using the method of variation of parameters, and requiring also
the free-boundary conditions (\ref{freecond}), we find:
$$
\vrh^{(\vsgo=0)}_1(\vph)=C_1\,\cos\vph+C_3+\Lambda_1\,\big(
3+2\lg(1+ \cos\vph)\big) \;.
$$
In addition, conditions (\ref{condfin}) yield:
\begin{eqnarray*}
\Lambda_1=-C_1\sin^2\frac{\psio}{2}\;,\qquad
C_3=C_1\left[1+\left(1+2\lg\left(2\sin^2\frac{\psio}{2}\right)
\right)\sin^2\frac{\psio}{2}\right]\;,&&\qquad
{\rm and}\\
2\left(1+\sin^2\frac{\psio}{2}+2\tan^2\frac{\psio}{2} \lg\sin^2
\frac{\psio}{2} \right)C_1=\frac{\Delta A}{\as}\;.&&
\end{eqnarray*}

When $\vsgo=1$, the particular solution of the differential
equation (\ref{lin1}) is again a constant, but the solution of the
homogeneous equation is not of the form (\ref{arlin}). If we solve
(\ref{lin1}) explicitly and use the free-boundary conditions
(\ref{freecond}), we arrive at:
$$
\vrh^{(\vsgo=1)}_1(\vph)=
C_1\,\cos\vph+C_3\,\big(\cos\vph\,\lg(1+\cos\vph)-1\big)-\Lambda_1\;.
$$
Finally, conditions (\ref{condfin}) now require:
\begin{eqnarray*}
C_1=-C_3\left(\lg\big(1-\cos\psio\big)-
\frac{\cos\psio}{1-\cos\psio}\right)\;,\qquad
\Lambda_1=-C_3\left(1+
\frac{\cos^2\psio}{1-\cos\psio}\right)\;,&&\qquad
{\rm and}\\
\cos^2\frac{\psio}{2}\cot^2\frac{\psio}{2} \;C_3=\frac{\Delta
A}{\as}\;.&&
\end{eqnarray*}

\begin{figure}
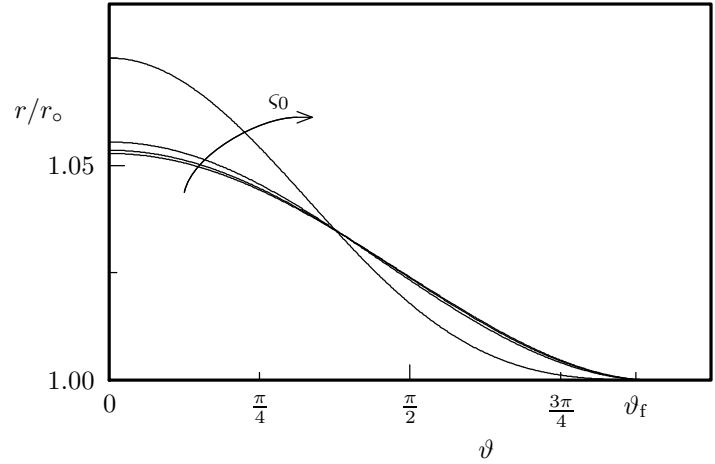

\input figure2.tex
\caption{Perturbed shapes of a vesicle hosting an inclusion, when
the area is slightly greater than the area corresponding to a
spherical equilibrium solution. The plots correspond to
$\psio=0.1\pi$, $\Delta A=0.05\,A_\circ$, and $\vsgo=0,1,2,3$ (the
arrow points towards increasing values of $\vsgo$).}
\label{figleg}
\end{figure}

\begin{figure}
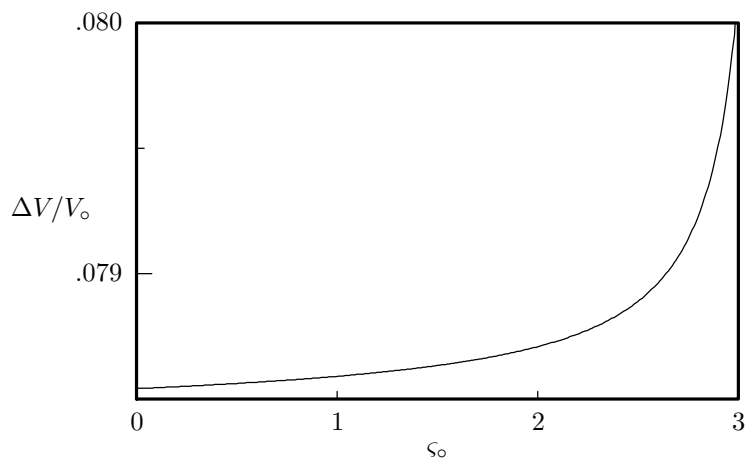

\input figure3.tex
\caption{Volume variation induced by a given area variation in a
permeable vesicle hosting an inclusion, as a function of the
reduced spontaneous curvature $\vsgo$. As in Figure \ref{figleg},
$\psio=0.1\pi$ and $\Delta A=0.05\,\as$.}
\label{arvol}
\end{figure}

Figure \ref{figleg} shows how the perturbed vesicle shape depends
on the spontaneous curvature for a prescribed area increase (5\%
with respect to the spherical value), while Figure \ref{arvol}
shows the volume variation induced by $\Delta A$.
\begin{itemize}
\item Since the inclusion is placed at $\vph=\vph_{\rm f}$, Figure
\ref{figleg} shows that in vesicles characterized by greater
spontaneous curvatures the shape modifications gather away from
the inclusion.
\item An area increase induces a shape perturbation $\vrh_1$ that
does not change sign all along the vesicle (we will find below
that this is not the case when also the volume is constrained).
\item Figure \ref{arvol} shows that the volume increase induced by
$\Delta A$ increases monotonically with $\Delta A$. However, the
remarkable increase in $\Delta V$ that shows up when $\vsgo\simeq
3$ is to be linked with the spontaneous-curvature driven budding
transition \cite{90wihe} that is close to occur. A detailed
analysis of the inclusion's influence on this transition can be
performed only by studying the nonlinear shape equation
(\ref{eulag}), and will be reported elsewhere \cite{04bina}.

\end{itemize}

\subsection{Small inclusions}

By using the asymptotic expansion (\ref{asexp}), it is possible to
show that in the small inclusion limit \hbox{$a\ll \sqrt{\as}$},
which implies $\psio\ll 1$ by virtue of (\ref{c1sph}), the
perturbed vesicle shape becomes independent of $\vsgo$:
\begin{equation}
r(\vph)=\rs\left(1+\frac{\Delta A}{\as}\;\frac{1+\cos\vph}{2}
\right)+O\left(\psio^2\lg\psio\,,\,\left(\frac{\Delta
A}{\as}\right)^2\right)\;.
\label{smallarea}
\end{equation}
Figure \ref{figleg} shows that when the spontaneous curvature is
small, the asymptotic expression (\ref{smallarea}) is more rapidly
approached. In fact, if we compute the volume variation associated
with the perturbed shape (\ref{smallarea}) we obtain:
$$
\frac{\Delta V}{\vs}=\frac{3}{2}\,\frac{\Delta
A}{\as}+O\left(\psio^2\lg\psio \,, \,\left(\frac{\Delta
A}{\as}\right)^2\right)\;.
$$
Figure \ref{arvol} confirms that the relative volume variation is
closer to the small limit prediction $\frac{3}{2}\Delta A/\as$
when $\vsgo$ is small.

\section{Impermeable vesicles \label{volume}}

We now focus on the solutions of the differential equation
(\ref{lin1}) that satisfy both area and volume constraints, when
these geometrical quantities are close to satisfy the spherical
condition (\ref{c2sph}). Leaving aside for the moment the cases
$\nu=0$ and $\nu=1$ (that turn out to be meaningless in the
impermeable case), the solution of (\ref{lin1}) is of the form
(\ref{form1}) with the following parameters to be determined:
$C_1$, $C_2$, $C_3$, $C_4$, $\Lambda_0$ \big(\emph{i.e.} $\nu$, by
virtue of (\ref{nu})\big), and the combination
$(\Lambda_1+\eta_1)$.

\subsection{Singular perturbations}

The boundary conditions (\ref{contcond}) and (\ref{freecond}) may
determine the four parameters $C_1$-$C_4$. A problem arises when
we try to determine $\nu$ and $(\Lambda_1+\eta_1)$ by using the
area and volume constraints. In fact, the $O(\eps)$ of the above
constraints reads as:
\begin{equation}
\int_0^{\vph_{\rm f}}\vrh_1(\vph)\sin\vph\,d\vph=\frac{\Delta
A}{4\pi\rs^2}\ ,\qquad\qquad \int_0^{\vph_{\rm
f}}\vrh_1(\vph)\sin\vph\,d\vph=\frac{\Delta V}{2\pi\rs^3}\;.
\label{nocons}
\end{equation}
It is clearly impossible to satisfy both constraints if
$$
\Delta A\neq\frac{2}{\rs}\,\Delta V\;.
$$
From the analytical point of view, the degeneracy of the
constraints (\ref{nocons}) stems from the fact that the first
variations of area and volume of a surface are linearly dependent
when computed in a spherical shape. In fact, if we perturb any
surface with constant mean curvature $H$ (the so-called Delaunay
surfaces \cite{DE1841, 95naok}), the area and volume of the
resulting surface satisfy \cite{89zhhe}
$$
\Delta A=2H\,\Delta V\,.
$$

The area and volume constraints become linearly independent only
when the second variations come into play. Thus, if we are willing
to perturb the assigned area and volume to an arbitrary $O(\eps)$,
we have to perturb the shape function $r(\vph)$ to
$O\big(\sqrt{\eps}\,\big)$, as we shall show next.

We begin by replacing (\ref{exp1}) by
\begin{equation}
r(\vph)=\rs\,
\big(1+\sqrt{\eps}\,\vrh_{\frac{1}{2}}(\vph)+\eps\,\vrh_1(\vph)
+o(\eps)\big)\;.
\label{exp1/2}
\end{equation}
The singular perturbation $\vrh_{\frac{1}{2}}$ is of the form
(\ref{form1}), with $\Lambda_1$ and $\eta_1$ replaced by their
half-order counterparts $\Lambda_{\frac{1}{2}}$ and
$\eta_{\frac{1}{2}}$. Correspondingly, the area and enclosed
volume of the resulting vesicle shape are given by:
\begin{align}
\nonumber\as+\epsilon\Delta A&=2\pi\int_0^{\vph_{\rm f}}
r\,\sqrt{r^2+r^{\prime
2}}\,\sin\vph\,d\vph\\
&=\as+4\pi\rs^2\sqrt{\eps}\int_0^{\vph_{\rm f}}
\!\!\!\vrh_{\frac{1}{2}}\sin\vph\,d\vph
+\pi\rs^2\eps\int_0^{\vph_{\rm f}}
\!\!\left(2\vrh^2_{\frac{1}{2}}+ \vrh^{\prime
2}_{\frac{1}{2}}+4\vrh_1\right) \sin\vph\,d\vph+o(\epsilon);
\label{a1}\\
\nonumber \vs+\epsilon\Delta V&=\frac{2\pi}{3}\int_0^{\vph_{\rm
f}} r^3(\vph)\sin\vph\,d\vph=\\
&=\vs+2\pi\rs^3\sqrt{\eps}\int_0^{\vph_{\rm f}}
\vrh_{\frac{1}{2}}\sin\vph\,d\vph +\pi\rs^3\eps\int_0^{\vph_{\rm
f}} \left(2\vrh^2_{\frac{1}{2}}+2\vrh_1\right)
\sin\vph\,d\vph+o(\epsilon)\;.
\label{v1}
\end{align}
Both (\ref{a1}) and (\ref{v1}) can now be satisfied, provided the
functions $\vrh_{\frac{1}{2}}$ and $\vrh_1$ are such that:
\begin{eqnarray}
&&\int_0^{\vph_{\rm f}} \vrh_{\frac{1}{2}}(\vph)\,\sin\vph\,d\vph
=0\;,\label{av1}\\
&&\int_0^{\vph_{\rm f}} \left[\vrh^{\prime
2}_{\frac{1}{2}}(\vph)-2\vrh^2_{\frac{1}{2}}(\vph)\right]
\sin\vph\,d\vph=\frac{\rs \Delta A-2\Delta V}{\pi\rs^3}\
,\qquad{\rm and}
\label{av2}\\
&&\int_0^{\vph_{\rm f}} \vrh_1(\vph) \sin\vph\,d\vph=\frac{\Delta
V}{2\pi\rs^3}-\int_0^{\vph_{\rm f}} \vrh^2_{\frac{1}{2}}(\vph)
\sin\vph\,d\vph\;.
\label{av3}
\end{eqnarray}
Equations (\ref{av1}) and (\ref{av2}) fix $\vrh_{\frac{1}{2}}$, as
we will show below; then, equation (\ref{av3}) allows to determine
also the now next-order correction $\vrh_1$.

Before entering in the detailed analysis of equations (\ref{av1})
and (\ref{av2}), the above result deserves some remarks.
\begin{itemize}
\item Equation (\ref{exp1/2}) underlines the most
striking effect of the impermeability constraint on the vesicle:
an $O(\eps)$ perturbation of the assigned geometrical values
induces an $O\big(\sqrt{\eps}\big)$ perturbation in the vesicle
shape.
\item Equation (\ref{av1}) is an eigenvalue equation. We have to
look for non-trivial (\emph{i.e.\/}, non-vanishing) perturbations
$\vrh_{\frac{1}{2}}$ that satisfy it. It will prove to admit a
countable infinity of independent solutions.
\item Equation (\ref{av2}) is quadratic in $\vrh_{\frac{1}{2}}$.
We will thus find two, rather than one, possible perturbed shapes
for any non-trivial solution of (\ref{av1}). An energy argument
will be needed to identify the preferred perturbation among the
double-infinity of possible choices at our disposal. Both
perturbed shapes arising from (\ref{av2}) will deserve notice,
since they display two qualitatively different vesicle reactions
to the perturbation.
\item The quadratic expression in the left-hand side of (\ref{av2})
forces the right-hand side combination $(\rs \Delta A-2\Delta V)$
to assume only non-negative values. This property is not peculiar
of vesicle theory: it reflects a classical isoperimetric
inequality. In fact, for any closed surface, the ratio $A^3/V^2$
is bounded from below by the value $36\pi$, which is attained only
by a sphere (see, \emph{e.g.\/}, \cite{51posz}, p.\ 8). Thus, for
example, it does not exist a closed surface with the same enclosed
volume of a sphere and smaller area.
\end{itemize}

\subsection{Multiplicity of stationary perturbed shapes}

\begin{figure}[htb]
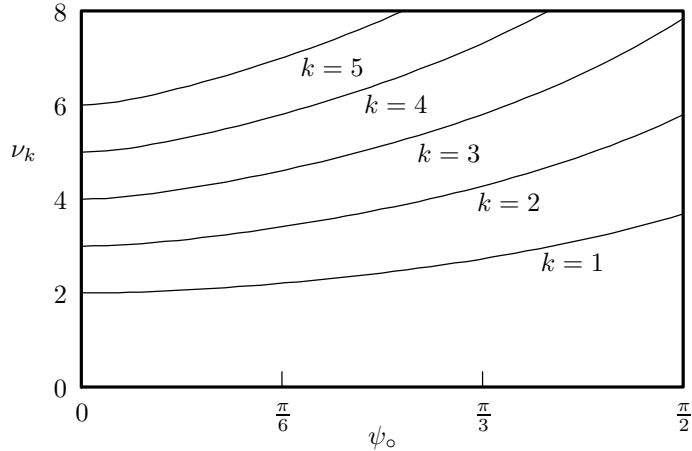

\input figure4.tex
\caption{Order $\nu$ of the Legendre functions entering in the
perturbation of an impermeable vesicle shape, as a function of the
inclusion apex angle $\psio$. The graphs display the smallest five
numerical solutions $\{\nu_k$, $k=1,\dots,5\}$ of the equation
(\ref{av1}).}
\label{nupsio}
\end{figure}

The free boundary conditions (\ref{freecond}) require that the
coefficients of the singular Legendre functions of the second kind
in (\ref{form1}) must be null: $C_2=C_4=0$. Afterwards, the
contact conditions (\ref{contcond}), which are linear in the shape
function, supply two relations that connect $C_1$, $C_3$, and
$(\Lambda_{\frac{1}{2}}+\eta_{\frac{1}{2}})$ in (\ref{form1}). As
a result, the leading perturbation $\vrh_{\frac{1}{2}}$ in
(\ref{exp1/2}) can be given the form:
\begin{align}
\vrh_{\frac{1}{2}}(\vph)&=C_3\,\big[\alpha(\nu,\psio)+\beta(\nu,\psio)
\cos\vph+P_\nu(\cos\vph)\big]\;,\qquad{\rm
with} \label{rho12} \\
\nonumber \alpha(\nu,\psio)&:=-\csc^2\psio\,\big( \nu\cos\psio \,
P_{\nu-1}(-\cos\psio) +(\sin^2\psio+\nu\cos^2\psio)
P_\nu(-\cos\psio)\big)\quad{\rm and}\\
\nonumber \beta(\nu,\psio)&:=-\nu\csc^2\psio\,\big(
P_{\nu-1}(-\cos\psio)+\cos\psio\,P_\nu(-\cos\psio)\big)\;.
\end{align}
In order to determine completely the function
$\vrh_{\frac{1}{2}}$, we have to use (\ref{av1}) and (\ref{av2})
to derive $\nu$ and $C_3$. Any non-trivial solution of (\ref{av1})
possesses $C_3\neq 0$. Thus, we can drop $C_3$ out from it, to
obtain an eigenvalue equation in $\nu$, depending only on $\psio$.
Figure \ref{nupsio} illustrates the numerical solutions of
(\ref{av1}). These solutions exhibit the following properties:
\begin{itemize}
\item The spontaneous curvature does not enter in (\ref{av1}).
Thus, the stationary shape modifications of an impermeable vesicle
do not depend on $\vsgo$, as they did in the permeable case (see
(\ref{nuso}) and Figure \ref{figleg}).
\item For any $\psio\in[0,\frac{\pi}{2}]$, there is a countable
infinity of values $\nu_k (\psio)$ satisfying (\ref{av1}).
\item For any $\psio\in[0,\frac{\pi}{2}]$, the solutions $\nu_k
(\psio)$ are symmetric with respect to $\nu=-\frac{1}{2}$.
However, since the symmetric solutions are identical (see
(\ref{8.2.1})), we can restrict our attention to solutions with
$\nu\geq -\frac{1}{2}$.
\item When $\psio\ll 1$ (small protein limit), the use of
(\ref{asexp}) allows to prove that $\nu_k$ tends to an integer
value for any $k$:
\begin{equation}
\nu_k(\psio)=(k+1)+\frac{k-1}{4}\;\psio^2+
o\left(\psio^2\right)\qquad {\rm for\ any}\quad k\in\filN\;.
\label{intg}
\end{equation}
The shape function thus approaches a linear combination of
Legendre functions of integer order, that is, Legendre
polynomials. Furthermore, both (\ref{intg}) and Figure
\ref{nupsio} show that only Legendre polynomials of order equal to
or greater than 2 come into play, while only low-order Legendre
polynomials ($P_0$ and $P_1$) entered the small-protein limit of
permeable vesicles (see (\ref{smallarea})). This yields more
drastic shape modifications in the incompressible case, since
Legendre polynomials are more and more oscillating as their order
increases. The expansion of the shape function (in the absence of
inclusions) in terms of Legendre polynomials was first used in
\cite{90wihe}.
\item For any $k\in\filN$, the functions $\nu_k(\psio)$ increase
monotonically with $\psio$, and do not intersect.
\item It is possible to prove by direct inspection that
$\nu=0$ and $\nu=1$ do not solve (\ref{av1}) for any value of
$\psio$.
\end{itemize}

Once we have identified the $\nu$-values that satisfy (\ref{av1}),
we can insert (\ref{rho12}) in (\ref{av2}) to determine $C_3$, the
only remaining free parameter in the singular perturbation
$\vrh_{\frac{1}{2}}$. We stress again that, being
$\vrh_{\frac{1}{2}}$ linear in $C_3$, this latter parameter enters
quadratically in (\ref{av2}). This fact, on the one hand fixes a
sign for the geometrical quantity $(\rs\Delta A-2\Delta V)$, and
on the other hand implies that, for any $k\in\filN$ with
$\nu=\nu_k(\psio)$, there are exactly two values of $C_3$ (one the
opposite of the other) that satisfy (\ref{av2}). Thus, for any
$k\in\filN$, there are two possible perturbed shapes:
$r_{k\pm}(\vph)=\rs\left(1\pm\sqrt{\eps}\,\vrh_{\frac{1}{2},k}(\vph)+
O(\eps)\right)$.

Only an energy estimate can help us in determining, for any
$\psio\in[0,\frac{\pi}{2}]$, both the value of $\nu_k$ and the
sign of $C_3$ that minimize the elastic energy. This will be the
aim of the remaining part of this section.

\subsection{Energy estimates}

\subsubsection{Ground state energy}

In order to identify which is energetically preferred among the
stationary perturbed shapes determined above, we will now compute
their elastic energy. When we insert (\ref{exp1/2}) in the
free-energy functional (\ref{fren}), and we make use of
(\ref{av1})-(\ref{av3}), we obtain:
\begin{align*}
\mathcal{F}_{k\pm}&=2\pi\kappa\int_0^{\vph_{\rm f}}\!\!
\big(H_{k\pm}-\sigma_0\big)^2\,r_{k\pm}\,\sqrt{r_{k\pm}^2+
r_{k\pm}^{\prime 2}}\,\sin\vph\,d\vph=\\
&=\kappa(\vsgo-1)^2\,\frac{\as}{\rs^2}+
\kappa\vsgo(\vsgo-1)\,\frac{\Delta A}{\rs^2}-
\kappa\vsgo\,\frac{\rs\Delta A-2\Delta V}{\rs^3}
+\mathcal{F}_k^{(2)}+o\big(\rs\Delta A,\Delta V\big)\;,
\end{align*}
where $\mathcal{F}_{k\pm}$ denotes the free energy of the $k$-th
solution of equation (\ref{av1}), with the positive or negative
sign for $C_3$, and
\begin{equation}
\mathcal{F}_k^{(2)}:=\frac{\pi}{2}\,\kappa \,\int_0^{\vph_{\rm
f}}\!\!\left[ \vrh_{\frac{1}{2},k}^{\prime
2}(\vph)\,\frac{\cos2\vph}{\sin^2\vph}+
\vrh_{\frac{1}{2},k}^{\prime\prime
2}(\vph)\right]\sin\vph\,d\vph\;
\label{bk}
\end{equation}
is the leading order that may determine $k$ and the sign of $C_3$.
However, $\mathcal{F}_k^{(2)}$ depends quadratically on
$\vrh_{\frac{1}{2},k}$, so that the sign of $C_3$ cancels out from
it. Thus, the second order expansion turns out to be able to
identify only the preferred value of $\nu_k$. A further term in
the free-energy expansion will be needed to complete the
determination of the free-energy absolute minimizer.

\begin{figure}[htb]
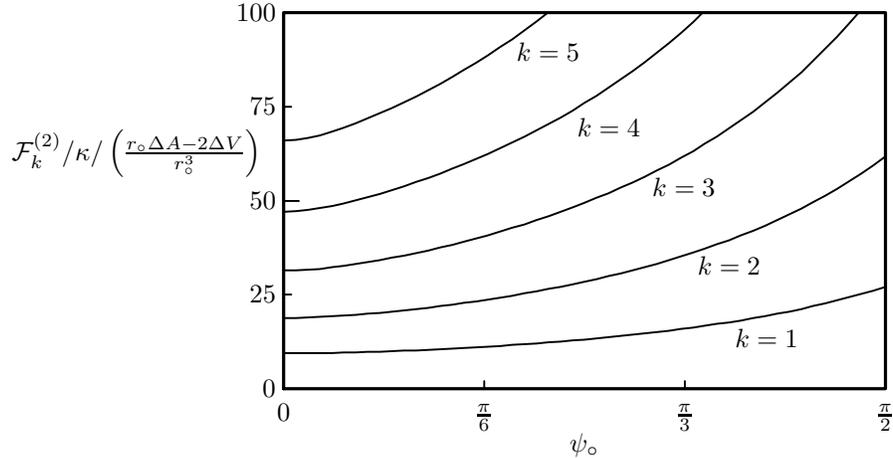

\input figure5.tex
\caption{Results of the numerical computation of the integral in
(\ref{bk}), when $\vrh_{\frac{1}{2},k}$ is given in (\ref{rho12}),
and $\nu$ is the $k$-th solution of equation (\ref{av1}). The
graphs display the results for $k=1,\dots,5$.}
\label{figbk}
\end{figure}

Figure \ref{figbk} shows how $\mathcal{F}_k^{(2)}$ depends on $k$
and $\psio$: the free energy increases when either of these
increase. In particular, Figure \ref{figbk} proves that the stable
perturbed shape for a quasi-spherical impermeable vesicle
corresponds to the solution with the smallest possible order of
the Legendre functions. This is not surprising from the physical
point of view, since Legendre functions wrinkle when their order
increases, and these oscillations increase the elastic energy.

\subsubsection{Pear-shaped or stomatocytes?}

\begin{figure}[htb]
\hfill
\includegraphics[height=4.1cm]{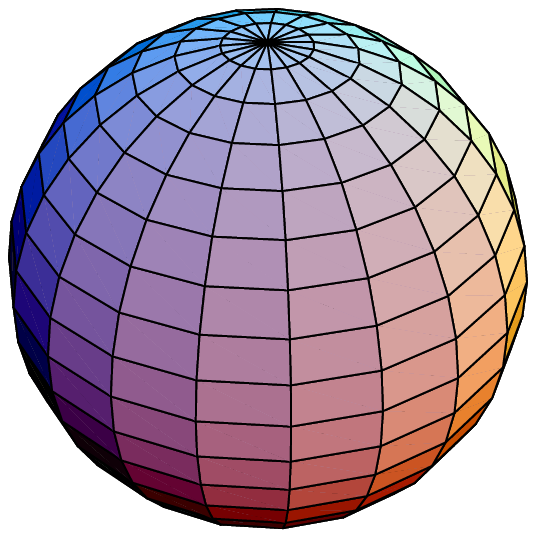}
\hfill
\includegraphics[height=4.35cm]{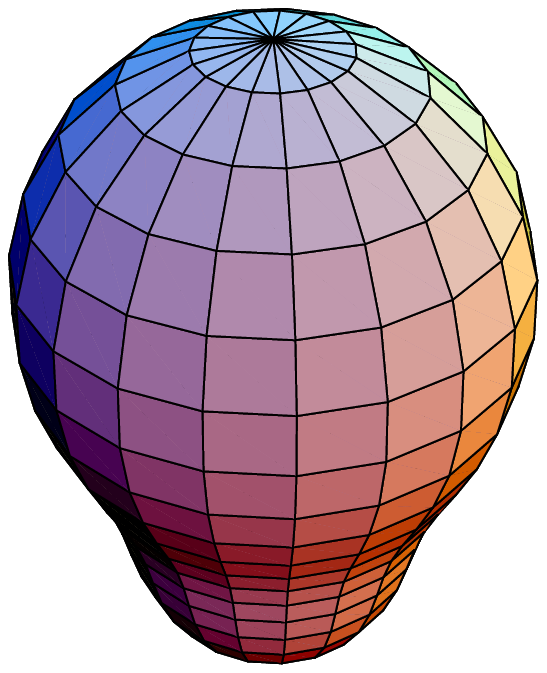}
\hfill
\includegraphics[height=3.5cm]{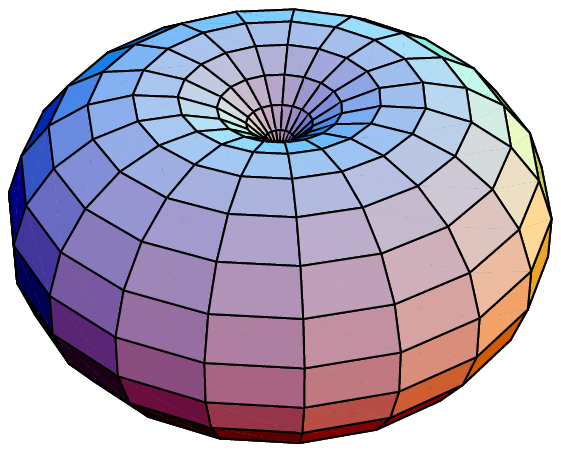}
\hfill\\
\null\hfill (a)\hskip4.2cm (b)\hskip4cm (c)\hfill\null
\caption{Perturbed shapes for a vesicle embedding an inclusion of
negligible size ($\psio\ll1$) when a 10\% increase in the vesicle
area is imposed (with respect to the value leading to a spherical
shape). Picture (a) shows the equilibrium shape of a permeable
vesicle, which is allowed to adapt its enclosed volume. Pictures
(b) and (c) refer to an impermeable vesicle when the positive or
negative sign for the parameter $C_3$ is chosen when solving
equation (\ref{av2}). The inclusion (not visible in the pictures)
sits in the bottom end of the vesicle.}
\label{pearstom}
\end{figure}

We still have to choose the preferred sign for the parameter
$C_3$. Figure \ref{pearstom} shows the deep, qualitative,
differences between permeable (a), and impermeable ((b) and (c))
stationary shapes that arise from the same parameter changes:
$\Delta A=\frac{1}{10}\as$ for all the shapes; $\Delta V$ is left
free in (a), while it is kept null in (b) and (c). Permeable
vesicles modify their enclosed volume in order to keep an
almost-spherical shape. On the contrary, impermeable vesicles move
towards pear-shaped or stomatocyte-like equilibrium shapes
\cite{91sebe}, depending on the sign of $C_3$.

\begin{figure}[htb]
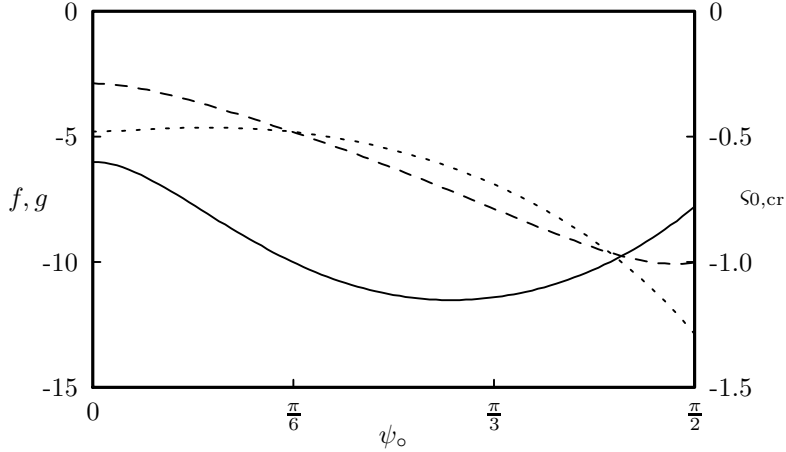

\input figure7.tex
\caption{Plots of the functions $f$ (dashed), $g$ (dotted line),
introduced in (\ref{ef3}), and the critical spontaneous curvature
$\varsigma_{0,\rm cr}$ (continuous line), defined in (\ref{socr}),
all as functions of the apex angle.}
\label{efgi}
\end{figure}

In order to compare the free energies of pear-shaped and
stomatocyte-like vesicles, we need to push further our free energy
expansion. The next order depends on the third power of
$\vrh_{\frac{1}{2}}$ and its derivatives: it is $O(\rs\Delta
A,\Delta V)^{3/2}$. By making use of the constraint requirements
(\ref{av1})-(\ref{av3}), it is possible to obtain:
$$
\mathcal{F}_{\rm (ps,st)}=\kappa(\vsgo-1)^2\,\frac{\as}{\rs^2}+
\kappa\vsgo(\vsgo-1)\,\frac{\Delta A}{\rs^2}-
\kappa\vsgo\,\frac{\rs\Delta A-2\Delta V}{\rs^3}
+\mathcal{F}^{(2)}+\mathcal{F}^{(3)}_{\rm (ps,st)}+o\big(\rs\Delta
A,\Delta V\big)^{3/2},
$$
where all the terms up to $\mathcal{F}^{(2)}$ do not depend on the
pear shaped \emph{vs.\/} stomatocyte choice, and
\begin{equation}
\mathcal{F}_{\rm (ps,st)}^{(3)}=\pm \left(\frac{\rs\Delta
A-2\Delta V}{\rs^3}\right)^{\frac{3}{2}}\big( f(\psio) + \vsgo\,
g(\psio) \big)\;.
\label{ef3}
\end{equation}
In (\ref{ef3}), the plus sign corresponds to the pear-shaped
vesicle, the minus sign describes a stomatocyte, and $\vsgo$
denotes as usual the reduced spontaneous curvature. Plots of the
functions $f$ and $g$ (whose explicit expressions can be found in
Appendix A2, equations (\ref{a22})-(\ref{a23})) are shown in
Figure \ref{efgi}. Both $f$ and $g$ are negative for all values of
$\psio$. Thus, for any positive value of the spontaneous
curvature, the pear-shaped vesicle is the absolute minimizer of
the free-energy, whereas the stomatocyte represents only a
relative minimum. A transition between the two stationary shapes
can be observed only when negative spontaneous curvatures are
induced; more precisely, the stomatocyte-like phase is preferred
if
\begin{equation}
\vsgo<-\frac{f(\psio)}{g(\psio)}=:\varsigma_{0,\rm cr}(\psio)\;.
\label{socr}
\end{equation}
Figure \ref{efgi} also shows how the critical value of the reduced
spontaneous curvature depends on the apex angle. In particular,
$-\frac{6}{5}<\varsigma_{0,\rm cr}(\psio)\leq -\frac{3}{5}$ for
all values of $\psio$.

\section{Concluding remarks}

In this paper we have analysed how the impermeability constraint
may induce singular stationary vesicle shapes, and how the
embedding of an inclusion may vary the vesicle topology, promoting
pear-shaped geometries that anticipate critical phenomena such as
budding or vesiculation \cite{91mifo,94fomi}.

Our analysis has been based on the linearization of the shape
equation close to a spherical shape. This approximation does not
allow to approach the aforementioned transitions, but in turn
yields analytical results which prove that the vesicle reaction to
a variation of the external parameters may not be analytical. In
Section 4 we have shown that an $O(\eps)$ relative variation in
the vesicle area may induce either an $O(\eps)$ or an
$O(\sqrt{\eps}\,)$ relative variation in the shape function,
depending on the permeability properties of the vesicle and the
aqueous solution that surrounds it.

The present analytical study is being currently completed by a
numerical study \cite{04bina}, which detects how the presence of
an embedded inclusion modifies the phase diagrams describing the
vesicle topology, and under which conditions it anticipates
budding phenomena.

\bigskip

\medskip\noindent{\bigbf Acknowledgements. }This work was made
possible by the Post-Doctoral Fellowship ``Mathematical Models for
Fluid Membranes'', supported by the Mathematical Department of the
\emph{Politecnico di Milano\/}. Some of the analytical
computations and figures presented in this paper were performed
with the aid of Mathematica$^{\circledR}$ 4.1, licence:
L4596-4499.

\renewcommand{\theequation}{A1.\arabic{equation}}
\setcounter{equation}{0}  
\section*{Appendix A1: Properties of Legendre functions}

The Legendre functions of the first and second kind (respectively
denoted as $P_\nu(s)$ and $Q_\nu(s)$) are the solutions of the
linear differential equation
$$
\big(1-s^2\big)\,y''(s)-2s\,y'(s)+\nu(\nu+1)\,y(s)=0\;.
$$
The Legendre functions of the first kind are regular when $s\to
1^-$ (with $P_\nu(1)=1$), while the Legendre functions of the
second kind are singular close to both $s=\pm 1$. The properties
we use in this paper are the following (see \cite{72abst}, \S 8.2,
and \S 8.5)

For any $\nu\in\filR$ and $s\in[-1,1]$,
\begin{align}
P_{-\frac{1}{2}-\nu}(s)&=P_{-\frac{1}{2}+\nu}(s)\;, \label{8.2.1}\\
P_{\nu+1}(s)&=\frac{2\nu+1}{\nu+1}\,s\,P_\nu(s)-\frac{\nu}{\nu+1}\,
P_{\nu-1}(s)\qquad (\nu\neq-1)\;,
\label{8.5.3} \\
\noalign{\medskip}
\big(1-s^2\big)\,P'_\nu(s)&=\nu\,P_{\nu-1}(s)-\nu\,s\,P_\nu(s)\;.
\label{8.5.4}
\end{align}
Equations (\ref{8.5.3}) and (\ref{8.5.4}) imply
\begin{equation}
\int P_\nu(\cos\vph)\,\sin\vph\,d\vph=-\int
P_\nu(s)\,ds=\frac{P_{\nu-1}(\cos\vph)-\cos\vph\,
P_\nu(\cos\vph)}{\nu+1}\qquad (\nu\neq-1)\;.
\label{intleg}
\end{equation}
For any $\nu\in\filR^+\setminus\filN$, the Legendre functions of
the first kind admit the following asymptotic expansion:
\begin{equation}
P_\nu(-1+\epsilon)=-\frac{\lg(\epsilon/2) +
2\gamma+\Psi(-\nu)+\Psi(\nu+1)}{\Gamma(-\nu)\,\Gamma(\nu+1)}+O(\epsilon\lg
\epsilon )\qquad{\rm as}\quad \epsilon\to 0^+\;,
\label{asexp}
\end{equation}
where $\Gamma$, $\Psi$, and $\gamma$ respectively denote the Euler
gamma function, the digamma function, and Euler's constant.

\renewcommand{\theequation}{A2.\arabic{equation}}
\setcounter{equation}{0}  
\section*{Appendix A2: Third order expansion of the free energy}

The derivation of the third-order term in the free-energy
expansion requires a third-order expansion of the shape function.
Thus, (\ref{exp1/2}) has to be replaced by:
$$
r(\vph)=\rs
\left(1+\sqrt{\eps}\,\vrh_{\frac{1}{2}}(\vph)+
\eps\,\vrh_1(\vph)+\eps^{\frac{3}{2}}\,\vrh_{\frac{3}{2}}(\vph)
+o\left(\eps^{\frac{3}{2}}\right)\right)\;.
$$
However, the higher-order terms $\vrh_1$ and $\vrh_{\frac{3}{2}}$
turn out to enter in the free-energy expansion only through
combinations that can be related to integrals of
$\vrh_{\frac{1}{2}}$ and its derivatives by making use of the area
and volume constraints, as it already happens in the second-order
expansion (see (\ref{av3})). More precisely, the third-order
expansion of the constraints yields
\begin{equation}
\int_0^{\vph_{\rm f}}
\vrh_{\frac{3}{2}}(\vph)\sin\vph\,d\vph+2
\int_0^{\vph_{\rm f}} \vrh_1(\vph)\,\vrh_{\frac{1}{2}}(\vph)
\sin\vph\,d\vph=-\frac{1}{3}\int_0^{\vph_{\rm f}}
\vrh^3_{\frac{1}{2}}(\vph) \sin\vph\,d\vph\;.
\label{a21}
\end{equation}
By using (\ref{a21}) it is long but straightforward to prove that
$$
\mathcal{F}_{\rm (ps,st)}=\kappa(\vsgo-1)^2\,\frac{\as}{\rs^2}+
\kappa\vsgo(\vsgo-1)\,\frac{\Delta A}{\rs^2}-
\kappa\vsgo\,\frac{\rs\Delta A-2\Delta V}{\rs^3}
+\mathcal{F}^{(2)}+\mathcal{F}^{(3)}+o\big(\rs\Delta A,\Delta
V\big)^{3/2}\;,
$$
with
$$
\mathcal{F}_{\rm (ps,st)}^{(3)}=F(\psio) + \vsgo\, G(\psio) \;,
$$
where
\begin{align*}
F(\psio) &=2\pi \int_0^{\vph_{\rm
f}}\left[\frac{\nu_k(1+\nu_k)}{3}
\vrh_{\frac{1}{2}}^3-\frac{\cos2\vph}{1-\cos2\vph}\vrh_{\frac{1}{2}}
\vrh_{\frac{1}{2}}^{\prime 2}+\frac{\cot\vph}{6}\,
\vrh_{\frac{1}{2}}^{\prime 3}-\frac{1}{2}\, \vrh_{\frac{1}{2}}
\vrh_{\frac{1}{2}}^{\prime\prime 2}\right]\sin\vph\,d\vph
\\
G(\psio) &=2\pi\int_0^{\vph_{\rm f}}\left[\vrh_{\frac{1}{2}}\,
\vrh_{\frac{1}{2}}^{\prime 2}-\frac{2}{3}\,\vrh_{\frac{1}{2}}^3+
\frac{1}{3}\vrh_{\frac{1}{2}}^{\prime 3}\cot\vph\right]\sin\vph\,
d\vph\;.
\end{align*}
The functions $f,g$ introduced in (\ref{ef3}) are related to $F,G$
through
\begin{align}
f(\psio) &=F(\psio)\;\sgn(C_3)\;\left(\frac{\rs\Delta A-2\Delta
V}{\rs^3}\right)^{-\frac{3}{2}}\;,
\label{a22}\\
g(\psio) &=G(\psio)\;\sgn(C_3)\;\left(\frac{\rs\Delta A-2\Delta
V}{\rs^3}\right)^{-\frac{3}{2}}\;.
\label{a23}
\end{align}

\goodbreak

\end{document}